\documentclass[11pt,a4paper]{article}
\usepackage{amsmath,amssymb}
\usepackage{epsfig,graphicx}
\usepackage{cite}

\begin{document}

\title{\bf A decay of the ultra-high-energy neutrino $\nu_e \to e^- W^+$ in a magnetic field 
and its influence on the shape \newline of the neutrino spectrum}

\author{A.~V.~Kuznetsov\footnote{{\bf e-mail}: avkuzn@uniyar.ac.ru},
N.~V.~Mikheev\footnote{{\bf e-mail}: mikheev@uniyar.ac.ru},
A.~V.~Serghienko\footnote{{\bf e-mail}: serghienko@mail.ru}
\\
\small{\em Yaroslavl State P.G.~Demidov University} \\
\small{\em Sovietskaya 14, 150000 Yaroslavl, Russian Federation}
}
\date{}

\maketitle

\begin{abstract}
The width of the neutrino decay into the electron and $W$ boson in a strong external 
magnetic field is obtained from the imaginary part of the neutrino self-energy. This result 
corrects the formulae existing in the literature. The mean free path of 
an ultra-high energy neutrino in a strong magnetic field is calculated. 
An energy cutoff for neutrinos propagating in a strong field is defined.
\end{abstract}

\section{Introduction}
\label{sec:Introduction}

\def\D{\mathrm{d}} 
\def\E{\mathrm{e}}
\def\I{\mathrm{i}}

The first and unique registration of extragalactic neutrinos from the supernova SN1987A 
explosion in the Large Magellanic Cloud, a satellite galaxy of our 
Milky Way, was undoubtedly an exciting achievement of neutrino astrophysics. 
The solution to the solar-neutrino puzzle in an experiment at the heavy-water detector installed
at the Sudbury Neutrino Observatory was one more important
result in this field. This experiment confirmed B. Pontecorvo's 
key idea concerning neutrino oscillations and, along with experiments that studied atmospheric 
and reactor neutrinos, thereby proved the existence of a nonzero neutrino mass and the
existence of mixing in the lepton sector, see e.g. \cite{Bilenky:2003,Mohapatra:2004} 
and the references cited therein. In this connection, the problem of studying the possible
effect of an active environment, including a strong magnetic field, on the dispersion properties 
of the neutrinos becomes quite important. 

An analysis of the effect of an external medium on neutrino properties relies on calculating the neutrino
self-energy operator $\Sigma(p)$, from which one can extract
the neutrino dispersion relation, and in part the imaginary 
part of the neutrino self-energy in medium, 
defining the width of the neutrino decay into the $W^+$ boson and a charged lepton, $\nu \to \ell^- W^+$. 
Here, we consider an electron as a charged lepton, 
but all the formulas are valid to the muon and $\tau$ lepton as well. 

A literature search reveals that calculations of the neutrino
dispersion relation in external magnetic fields have a long
history \cite{McKeon:1981, Borisov:1985, Erdas:1990, Erdas:2003, 
Kuznetsov:2006, Kuznetsov:2007, Bhattacharya:2009, Erdas:2009}. To compare the
different results we analyse the neutrino self-energy operator
$\Sigma (p)$ that is defined in terms of the invariant amplitude for
the transition $\nu_e \to \nu_e$ by the relation
\begin{equation}
{\cal M} (\nu_e \to \nu_e) = - \left[\bar\nu_e (p) \,\Sigma (p) \, \nu_e (p) \right] = 
- \textrm{Tr} \left[\Sigma (p) \, \rho (p) \right],
\label{eq:sigma_def}
\end{equation}
where $p = (E, \, {\bf p})$ is the neutrino four-momentum, 
$\rho (p) = \nu_e (p) \bar\nu_e (p)$ is the neutrino density matrix. 
On the other hand, the additional energy $\Delta E$ acquired 
by a neutrino in an external magnetic field is defined via the invariant 
amplitude~(\ref{eq:sigma_def}) as follows: 

\begin{equation}
\Delta E = - \frac{1}{2 \, E} \, {\cal M} (\nu_e \to \nu_e) \,.
\label{eq:Delta_E_def}
\end{equation}

\begin{figure}[htb]
\centering
\includegraphics{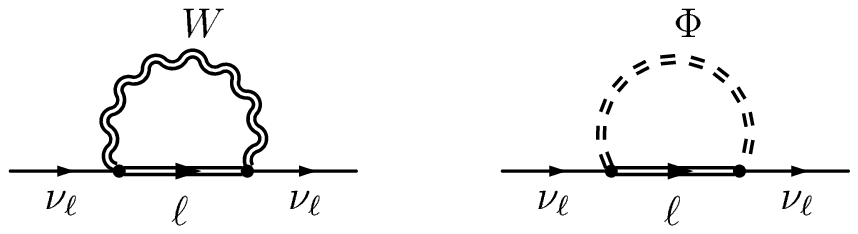}
\caption{Feynman diagrams representing the magnetic-field-induced
contribution to the neutrino self-energy operator
in the Feynman gauge. Double lines correspond to the
exact propagators for the charged lepton, the $W$ boson,
and the nonphysical scalar charged $\Phi$ boson in an external
magnetic field.}
\label{fig:Feynman}
\end{figure}

The $\cal S$ matrix element for the transition $\nu_e \to \nu_e$
corresponds to the Feynman diagrams shown in Fig.~\ref{fig:Feynman}
where double lines denote exact propagators in the presence of an
external magnetic field. A detailed description of the calculational
techniques for the neutrino
self-energy operator $\Sigma(p)$ in external electromagnetic fields
can be found e.g. in Ref.~\cite{Erdas:1990}, see 
also~\cite{Kuznetsov:2006,Kuznetsov:2007,Kuznetsov:2003}.
The relevant $\cal S$-matrix element can be used to
deduce, in a standard way, the invariant amplitude~(\ref{eq:sigma_def}),
whence the neutrino self-energy operator takes the form
\begin{eqnarray}
\Sigma (p) = - \frac{\mathrm{i} \,g^2}{2} \biggl[ \gamma^\alpha \, L \, 
J^{(W)}_{\alpha \beta} (p) \, \gamma^\beta \, L 
+ \frac{1}{m_W^2}
\left(m_e R - m_\nu L \right) J^{(\Phi)} (p) 
\left(m_e L - m_\nu R \right) \biggr] .
\label{eq:sigma1}
\end{eqnarray}
Here, $g$ is the Standard Model electroweak coupling
constant; $\gamma_\alpha$ are the Dirac matrices; 
$L = (1-\gamma_5)/2$ and $R = (1+\gamma_5)/2$
are, respectively, the left- and the right-hand projection
operator. The integrals introduced in~(\ref{eq:sigma1}) have the form
\begin{eqnarray}
J^{(W)}_{\alpha \beta} (p) &=& \int \, \frac{\mathrm{d}^4 q}{(2 \pi)^4} \,S (q) \, 
G^{(W)}_{\beta \alpha} (q-p) \,, 
\nonumber\\
J^{(\Phi)} (p) &=& \int \, \frac{\mathrm{d}^4 q}{(2 \pi)^4} \,S (q) \, D^{(\Phi)} (q-p) \,,
\label{eq:J_phi_def}
\end{eqnarray}
where $S (q)$, $G^{(W)}_{\beta \alpha} (q-p)$ and $D^{(\Phi)} (q-p)$ 
are the Fourier transforms of the translation-invariant
parts of the propagators for the electron, the $W^-$ boson, and the charged scalar
$\Phi$ boson, respectively. We note that the quantity $m_\nu$
in~(\ref{eq:sigma1}) is in general the nondiagonal Dirac neutrino 
mass matrix with allowance for mixing in the lepton sector.
By this means the flavor non-conserving decays $\nu_e \to \mu W, \tau W$ are 
also possible which are suppressed, however, by the very small parameter 
$\sim (m_\nu/m_W)^2 \lesssim 10^{-22}$. 


The general Lorentz structure of the operator $\Sigma (p)$
in a magnetic field, defined in Eq.~(\ref{eq:sigma1}), 
can be represented in the form~\cite{Kuznetsov:2007}
\begin{eqnarray}
\Sigma(p)&=&
\left[{\cal A}_L\,(p\gamma) + {\cal B}_L \,(p \gamma)_{\|} + 
{\cal C}_L \, (p \tilde \varphi \gamma) \right]\, L 
\nonumber\\
&+& \left[ {\cal A}_R\,(p\gamma) + {\cal B}_R \,(p \gamma)_{\|} + 
{\cal C}_R \, (p \tilde \varphi \gamma) 
\right]\, R 
+ m_\nu \, \left[{\cal K}_1 + \mathrm{i} \, {\cal K}_2 \left(\gamma \varphi \gamma \right) \right]
\,. 
\label{eq:sigma2}
\end{eqnarray}
The Lorentz indices of four-vectors and tensors
within parentheses are contracted consecutively, e.g. 
$(p\varphi\gamma) = p^{\alpha}\varphi_{\alpha\beta}\gamma^{\beta}$.
Further, $\varphi$ is the dimensionless tensor of the 
electromagnetic field, normalized to the external $B$-field, whereas
$\tilde\varphi$ is its dual,
\begin{eqnarray}
\varphi_{\alpha \beta} =  \frac{F_{\alpha \beta}}{B}\,,
\qquad 
{\tilde \varphi}_{\alpha \beta}=
\frac{1}{2} \, \varepsilon_{\alpha \beta \mu \nu} \varphi^{\mu \nu}\,. 
\label{eq:phi}
\end{eqnarray}
Finally, in the frame where only an external magnetic field $\bf B$ is
present, we take the spatial 3-axis to be directed along $\bf B$.
Four-vectors with the indices $\bot$ and $\|$ belong to the Euclidean
$\{1, 2\}$-subspace and the Minkowski $\{0, 3\}$-subspace,
correspondingly. For example, $p_\bot=(0,p_1,p_2,0)$ and
$p_\|=(p_0,0,0,p_3)$.  For any four-vectors $P$ and $Q$ we use the
notation
\begin{eqnarray}
(PQ)_{\|} &=& (P \,\tilde \varphi \tilde \varphi \,Q) = 
P_0 Q_0 - P_3 Q_3\,,
\nonumber\\
(PQ)_{\perp} &=& (P \,\varphi \varphi \,Q) =  P_1 Q_1 + P_2 Q_2\,, 
\nonumber\\
(PQ) &=& (PQ)_{\|} - (PQ)_{\perp}\,.
\label{eq:notation}
\end{eqnarray}

The coefficients ${\cal A}_R$, ${\cal B}_R$, ${\cal C}_R$, and ${\cal K}_{1,2}$ in~(\ref{eq:sigma2}) 
stem from the Feynman diagram involving the scalar $\Phi$ boson,
while the coefficients ${\cal A}_L$, ${\cal B}_L$, and ${\cal C}_L$ contain the
contributions from both diagrams. 
We note that the
coefficients ${\cal A}_L$, ${\cal A}_R$, and ${\cal K}_1$ 
in~(\ref{eq:sigma2}) contain an
ultraviolet divergence which is removed by the vacuum renormalization
of the neutrino wave function and mass.

Using Eqs.~(\ref{eq:sigma_def}), (\ref{eq:Delta_E_def}) and~(\ref{eq:sigma2}),
the neutrino additional energy $\Delta E$ in an external magnetic field
can be written in the form:
\begin{eqnarray}
\Delta E &=& {\cal B}_L \, \frac{p_\|^2}{2 E} \left[ 1 - ({\bf s} {\bf v}) \right]  
+ {\cal B}_R \, \frac{p_\|^2}{2 E} \left[1 + ({\bf s} {\bf v}) \right]  
\nonumber\\
&-& \frac{m_\nu}{2} \left[ {\cal C}_L - {\cal C}_R + 4 {\cal K}_2 
- ({\cal B}_L - {\cal B}_R )\, ({\bf b} {\bf v})
\right] \left[({\bf s} {\bf b}_t ) + \frac{m_\nu}{E} \, ({\bf s} {\bf b}_{\ell} ) \right]
\nonumber\\
&+& \frac{m_\nu^2}{2 E} \left( {\cal A}_L + {\cal A}_R + 2 {\cal K}_1 \right) \,,
\label{eq:Delta_E_2}
\end{eqnarray}
where ${\bf v} = {\bf p}/E$ is the neutrino velocity vector, 
${\bf s}$ is the unit vector of the doubled neutrino spin, 
${\bf b}$ is the unit vector along the magnetic field direction, and 
${\bf b}_{t,\ell}$ are its transversal and longitudinal components with respect 
to the neurino momentum, ${\bf b} = {\bf b}_t + {\bf b}_{\ell}$. 

In the previous papers, the neutrino self-energy operator~(\ref{eq:sigma1}) 
was calculated in different regions of values of the physical parameters, 
however, the list of these considered regions appears not to be comprehensive.
Namely, the investigated limiting cases were the following: 
\begin{itemize}
\item[i)] 
a weak field case 
($e B \ll m_e^2$)~\cite{Erdas:1990, Kuznetsov:2006}; 
\item[ii)]                 
a case of a moderately strong field ($m_e^2 \ll e B \ll m_W^2$) and limited region of a neutrino transverse momentum 
with respect to the magnetic field ($p_{\perp} \ll m_W$)~\cite{Kuznetsov:2006}; 
\item[iii)] 
the situation where the neutrino transverse momentum
$p_{\perp}$ is rather high, for example, $p_{\perp} \gtrsim m_W$ or $p_{\perp} \gg m_W$, 
while the magnetic field strength is not too high, 
$e B \ll m_e^2$, which corresponds
to the crossed-field 
approximation~\cite{Borisov:1985, Erdas:2003, 
Kuznetsov:2007, Bhattacharya:2009}. 
\end{itemize}

There is yet another region of values of the physical parameters
that requires a dedicated analysis. We mean here 
the case of the high neutrino transverse momentum,
when the magnetic field strength is also rather high, 
thus, the crossed-field approximation is not valid. 

This region of parameter values is of importance in connection with
problems of the physics of magnetars, the pulsars with
superstrong surface magnetic fields ($B_s \sim 10^{15}$ G). 
In particular, the possibility of detecting cosmic
neutrinos of ultrahigh energy, $\sim$ 1 PeV or even higher, from magnetars 
is widely discussed (see, for example,~\cite{Zhang:2003,Luo:2005,Ioka:2005}). 
It looks reasonable that the process of emission of
neutrinos having such energies cannot be described
adequately without taking into account their interaction
with a strong magnetic field of a magnetar.

This talk is based on our recent paper~\cite{Kuznetsov:2010}. 

\section{Charged-lepton, $W$- and $\Phi$-boson propagators in a
magnetic field}                                
\label{sec:propagators}

The Fourier transforms of the translation-invariant parts of the
exact propagators in an external magnetic field, entering 
into expressions~(\ref{eq:J_phi_def}) can be presented in the
Fock proper-time formalism in the following form, see e.g.~\cite{Schwinger:1951,Erdas:1990}.  
The lepton propagator is
\begin{eqnarray}
S (q) = \int_0^{\infty}\!\! 
\frac{\mathrm{d} s}{\cos \beta s}\, \, \mathrm{e}^{-\mathrm{i} \Omega_e} \,
\biggl\{\left[(q \gamma)_{\|} + m_\ell\right] \left[
\cos \beta s - \frac{ (\gamma \varphi \gamma)}{2} \,
\sin \beta s \right]-
\frac{(q \gamma)_{\perp}}{\cos \beta s}\biggr\},
\label{eq:S(q)}
\end{eqnarray}
where $\beta = e B$ and $m_e$ is the electron mass. 

Similarly, the
$W$-boson propagator can be written as 
\begin{eqnarray}
G_{\rho \sigma} (q) = - \int_0^{\infty} 
\frac{\mathrm{d} s}{\cos \beta s} \, \mathrm{e}^{-\mathrm{i} \Omega_W} \,
\biggl[
(\tilde\varphi \tilde\varphi)_{\rho \sigma} 
- (\varphi \varphi)_{\rho \sigma} \, \cos \, 2 \beta s
- \varphi_{\rho \sigma} \, \sin \, 2 \beta s \biggr] .
\label{eq:G(q)}
\end{eqnarray}

And finally, for the $\Phi$-boson propagator one obtains
\begin{eqnarray}
D^{(\Phi)} (q) = \int_0^{\infty} \mathrm{d} s \, \mathrm{e}^{-\mathrm{i} \Omega_W} \,,
\label{eq:D(q)}
\end{eqnarray}
where we have chosen the Feynman gauge for the
$W$ and $\Phi$ bosons and have introduced the notation
($j = e, W$)
\begin{equation}
\Omega_j = s \left(m_j^2- q_{\|}^2 \right) 
+ \frac{\tan \beta s}{\beta}\,q_{\perp}^2 \,.
\label{eq:Omega}
\end{equation}
%

\section{The neutrino decay $\nu \to e^- W^+$ in an external 
electromagnetic field}
\label{sec:decay}

The probability of the neutrino decay $\nu \to e^- W^+$ 
in an external electromagnetic field is one of the most
interesting results that can be extracted from the neutrino
self-energy operator. 
This probability can be expressed in terms of the imaginary part of the 
amplitude~(\ref{eq:sigma_def}) with the neutrino self-energy 
operator~(\ref{eq:sigma2}). 

For simplicity, hereafter we neglect the neutrino mass $m_\nu$, taking 
the density matrix of the left-handed neutrino as $\rho (p) = (p\gamma) \, L$. 
One obtains:
\begin{eqnarray}
w (\nu \to e^- W^+) = \frac{1}{E}\, \textrm{Im} \, {\cal M}(\nu_e \to \nu_e) 
= - \frac{1}{E}\, \textrm{Im} \, \textrm{Tr} \left[\Sigma (p) \, (p\gamma) \, L \right]
= - 2 \, \frac{p_{\perp}^2}{E}\, \textrm{Im} \, {\cal B}_L \,.
\label{eq:width}
\end{eqnarray}

An analysis of the neutrino decay $\nu \to e^- W^+$ in
an external field is of interest only at ultrahigh neutrino
energies. 

In all previous papers the neutrino decay width in an external electromagnetic field was 
calculated in the crossed field approximation, 
in which case the width is expressed in terms 
of the dynamical field parameter $\chi$ 
and the lepton mass parameter $\lambda$: 
\begin{eqnarray}
\chi = \frac{e (p F F p)^{1/2}}{m_W^3}\,,  
\qquad \lambda = \frac{m_e^2}{m_W^2}\,.
\label{eq:def_chi}
\end{eqnarray}
The particular case of a crossed field 
is in fact more general than it may seem at first glance. 
Really, the situation is possible when the 
field dynamical parameter 
$\chi$ of the relativistic particle 
propagating in a relatively weak electromagnetic field, $F < B_e$ 
(where $F$ means the electric and/or magnetic field strength, 
$B_e = m_e^2/e \simeq 4.41 \times 10^{13}\,\textrm{G}$ is the critical
field value), could appear rather high. 
In this case the field in the particle rest 
frame can exceed essentially the critical 
value and is very close to the crossed field. 
Even in a magnetic field whose strength is much greater than the 
critical value, the result obtained in a crossed field will correctly 
describe the leading contribution to the probability of a process in a 
pure magnetic field, provided that $\chi \gg B/B_e$. 
In the frame where the field is pure magnetic one, the dynamical field parameter takes the form:
\begin{eqnarray}
\chi = \frac{e B \ p_{\perp}}{m_W^3}\,.
\label{eq:def_chi2}
\end{eqnarray}

A general expression for the decay width can be written in this case in the 
form~\cite{Kuznetsov:2007}
\begin{eqnarray}
w (\nu_e &\to& e^- W^+) = \frac{\sqrt{2} \,G_{\rm F} \, m_W^4}{12 \sqrt{3} \, \pi^2 \, E}  
\int\limits_0^{1} \frac{\mathrm{d} z}{z (1-z)^2} \, K_{2/3} (U) 
\nonumber\\
&\times& \, [z + \lambda \,(1 - z)] \, 
[2 (1 + z) (2 + z) + \lambda \,(1 - z) (2 - z)] \,,
\label{eq:w_our07}
\end{eqnarray}
where $K_{2/3} (U)$ is the modified Bessel function, with the argument: 
\begin{eqnarray}
U = \frac{2}{3 \, \chi} \, \frac{[z + \lambda \,(1 - z)]^{3/2}}{z (1-z)} \,.
\label{eq:Bessel_arg}
\end{eqnarray}

Taking in Eq.(\ref{eq:w_our07}) the limit $\chi, \lambda \ll 1$, one obtains the result which 
can be written in terms of the only modified dynamical field parameter 
\begin{eqnarray}
\xi = \frac{\chi}{\sqrt{\lambda}} = \frac{e B \ p_{\perp}}{m_e \, m_W^2} \,. 
\label{eq:xi}
\end{eqnarray}
The range for the $\xi$ parameter appears to be rather large, $0 < \xi \ll 1/\sqrt{\lambda} = m_W/m_e$. For the electron, it is: $0 < \xi \ll 1.6 \times 10^5$. If one replaces $m_e \to m_\tau$, then the range  
is not too wide, $0 < \xi \ll 45$. 

The decay width takes the form 
\begin{eqnarray}
w (\nu \to e^- W^+) = \frac{\sqrt{2} \,G_{\rm F}}{3 \pi} \,
\frac{(e B \ p_{\perp})^2}{m_W^2 \, E} \, 
\left( 1 + \frac{\sqrt{3}}{\xi} \right) \exp \left( - \frac{\sqrt{3}}{\xi} \right) \,.
\label{eq:w_Erdas03}
\end{eqnarray}

The formula~(\ref{eq:w_Erdas03}) should be compared with 
the results of Refs.~\cite{Borisov:1985,Erdas:2003,Bhattacharya:2009}. 
It should be mentioned that the decay width $w$ defined in Refs.\cite{Borisov:1985,Kuznetsov:2007} 
is the same, in the natural system of units, than the absorption coefficient 
$\alpha$ \cite{Erdas:2003} and the damping rate of the neutrino 
$\gamma$ \cite{Bhattacharya:2009}. 
One can see that the absorption coefficient 
$\alpha$ presented in Eq.(25) of Ref.~\cite{Erdas:2003} looks very similar 
to our Eq.~(\ref{eq:w_Erdas03}). 
However, the angular dependence in our formulas is quite different: 
instead of the factor $p_{\perp}^2/E = E \sin^2 \theta$ standing in our 
Eq.~(\ref{eq:w_Erdas03}), there is the factor $p_{\perp} = E \sin \theta$ 
in Eq.(25) of Ref.~\cite{Erdas:2003}. 

On the other hand, one can see that our result~(\ref{eq:w_Erdas03}) 
surely contradicts the Eq.~(58) of Ref.~\cite{Bhattacharya:2009}, where an attempt was made 
of reinvestigation of the process $\nu \to e^- W^+$ in the crossed field approximation. 
The difference is the most essential at small values of $\xi$, where the result 
of Ref.~\cite{Bhattacharya:2009} appears to be strongly underestimated. 

In the earlier paper by Borisov et al.~\cite{Borisov:1985} the calculations 
of the process $\nu \to e^- W^+$ width were performed 
in the two limiting cases of the small and large values of the parameter $\chi$.  
In the limit $\chi^2 \ll \lambda$ (i.e. $\xi \ll 1$) their result can be presented in the form
\begin{eqnarray}
w = \frac{\sqrt{2} \ G_{\rm F}}{\sqrt{3} \ \pi} \ m_e \ eB \sin \theta 
\exp \left( - \sqrt{3} \ \frac{m_e m_W^2}{e B \ p_{\perp}} \right) \,,
\label{eq:w_Bor_small_chi}
\end{eqnarray}
and can be reproduced from the general formula~(\ref{eq:w_Erdas03}).

On the other hand, in the limit $\chi \gg 1$ ($\xi \gg 1/\sqrt{\lambda}$) the 
result of Ref.~\cite{Borisov:1985} can be written as
\begin{eqnarray}
w = \frac{\sqrt{3} \ G_{\rm F}}{\sqrt{2} \ \pi} \ m_W \ e B \sin \theta \,,
\label{eq:w_Bor_big_chi}
\end{eqnarray}
and can be reproduced from our more general formula~(\ref{eq:w_our07}).

A problem of the decay $\nu \to e^- W^+$ has a physical meaning only in the fields of the pulsar type, 
where the field strength is of order of the critical value $\sim 10^{13}$ G. 
The above formulas for the probability except for Eq.~(\ref{eq:w_Bor_big_chi}) are applicable 
for relatively weak fields only, $B \ll 10^{13}$ G. Taking into account the discovery of magnetars which 
are the neutron stars with the fields $\sim 10^{14} - 10^{15}$ G, it is interesting to 
calculate the probability of the process $\nu \to e^- W^+$ in such fields where the crossed-field approximation 
is inapplicable. 

Thus, we will use the following hierarchy of the physical parameters: 
$p_{\perp}^2 \gg m_W^2 \gg e B \gg m_e^2$. A general expression for the process $\nu \to e^- W^+$ 
probability can be obtained by the substitution of Eq.~(\ref{eq:sigma1}) into Eq.~(\ref{eq:width}) 
with taking account of Eqs.~(\ref{eq:S(q)}) - (\ref{eq:D(q)}). After calculations which are not difficult 
but rather cumbersome, the process width can be expressed via the function $\Phi (\eta)$ depending on the one parameter $\eta$ only: 
\begin{eqnarray}
\eta = \frac{4 \ e B p_{\perp}^2}{m_W^4} \,.
\label{eq:def_eta}
\end{eqnarray}

The process width can be presented in the form 
\begin{eqnarray}
w (\nu \to e^- W^+) = \frac{G_{\rm F} \, (e B)^{3/2} \ p_{\perp}}{\pi \sqrt{2 \pi} \, E} \;
\Phi (\eta) \,,
\label{eq:w_our_gen}
\end{eqnarray}
where
\begin{eqnarray}
\Phi (\eta) = \frac{1}{\eta} \int\limits_0^{\infty} \frac{\mathrm{d} y}{y^{1/2}} 
\frac{(\tanh y)^{1/2}}{(\sinh y)^2} 
\frac{(\sinh y)^2 - y \, \tanh y}{(y - \tanh y)^{3/2}}
\exp \left[ - \frac{y \, \tanh y}{\eta (y - \tanh y)} \right] .
\label{eq:phy(eta)_int}
\end{eqnarray}
We stress that we have obtained this formula neglecting the electron mass 
as the smallest parameter in the hierarchy used.

The formulas~(\ref{eq:w_our_gen}), (\ref{eq:phy(eta)_int}) present our main results and are valid 
in a wide region of the parameter $\eta$ values, $0 < \eta \ll m_W^2/(e B)$. 
The function $\Phi (\eta)$ is essentially simplified at large and small values of the argument. 

In the limit $\eta \gg 1$ one obtains:
\begin{eqnarray}
\Phi (\eta \gg 1) \simeq \frac{1}{3} \, \sqrt{\pi (\eta - 0.3)} \,, 
\label{eq:phy_large_eta}
\end{eqnarray}
and the error is less than 1 \% for $\eta > 10$. 

The formulas~(\ref{eq:w_our_gen}), (\ref{eq:phy_large_eta}) reproduce the probability~(\ref{eq:w_Erdas03}),  where the limit $\xi \gg 1$ should be taken.  

In the other limit $\eta \ll 1$ one obtains
\begin{eqnarray}
\Phi (\eta \ll 1) \simeq \exp \left( - \frac{1}{\eta} \right) 
\left(1 - \frac{1}{2} \, \eta + \frac{3}{4} \, \eta^2 \right) \, 
\label{eq:phy_small_eta}
\end{eqnarray}
and the error is less than 1 \% for $\eta < 0.5$. 

The formulas obtained allow to establish an upper limit on the energy spectrum of neutrinos 
propagating in a strong magnetic field. 
Let us take the typical size $R$ of the region with the strong magnetic field as $R \sim $ 10 km. 
If the neutrino mean free path $\lambda = 1/w$ is much less than the field size, $\lambda \ll R$, all the 
neutrinos are decaying inside such the field. For $\lambda =$ 1 km $\ll R$, we can find the cutoff energies $E_c$
for the neutrino spectrum, depending on the magnetic field strength. 
This dependence is calculated numerically and is shown in Fig.~\ref{fig:cutoff}. There are two parameter 
regions where the dependence can be essentially simplified, as follows:


\begin{figure}[htb]
\centering
\includegraphics[scale=0.8]{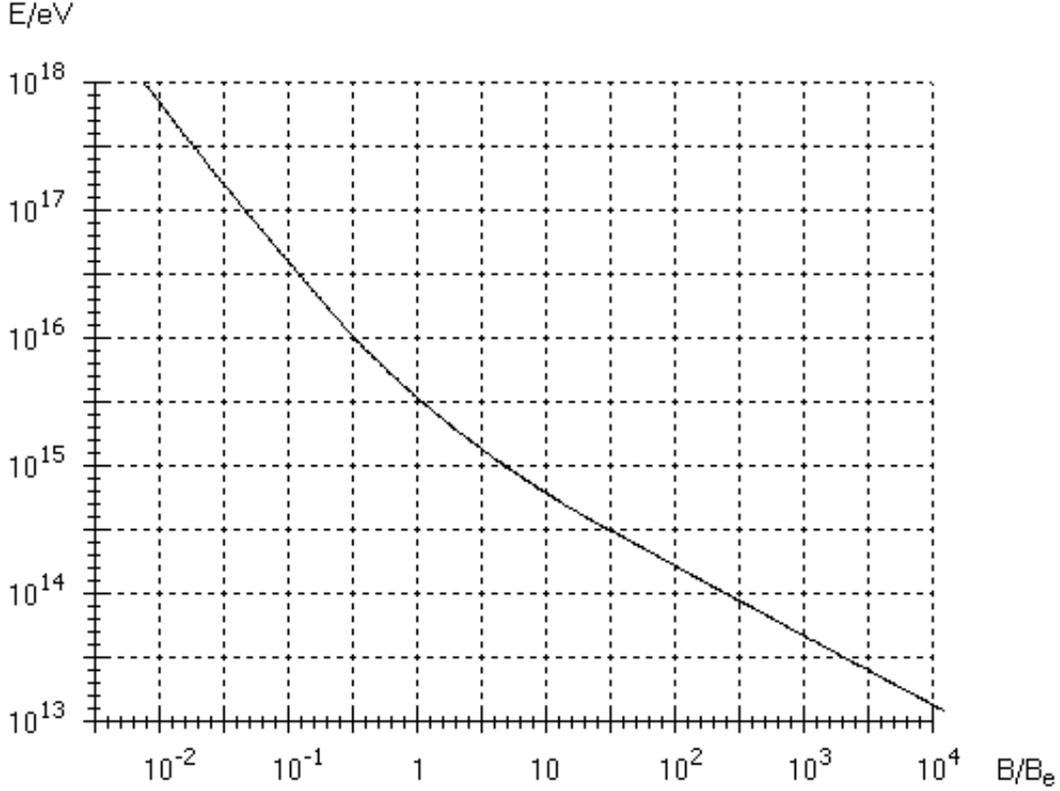}
\caption{The dependence of the neutrino cutoff energy $E_c$ on the magnetic field strength for the fixed neutrino mean free path value, $\lambda =$ 1 km.}
\label{fig:cutoff}
\end{figure}


i) for relatively weak field, $B \simeq 0.1 B_e \simeq 4 \times 10^{12}$ G, 
the neutrino mean free path can be obtained from Eq.~(\ref{eq:w_Bor_small_chi}):
\begin{eqnarray}
\lambda \simeq \frac{4.9 \, \textrm{m}}{B_{0.1} \, \sin \theta} \, 
\exp \left( \frac{219}{B_{0.1} \, E_{15} \, \sin \theta} \right) ,
\label{eq:lambda_w}
\end{eqnarray}
where $B_{0.1} = B/ (0.1 B_e)$, $E_{15} = E/(10^{15} \textrm{eV})$, and the cutoff energy corresponding to 
$\lambda =$ 1 km, at $B_{0.1} = 1$, $\theta = \pi/2$, is 
\begin{eqnarray}
E_c \simeq 0.4 \times 10^{17} \textrm{eV} \,;
\label{eq:E_c_w}
\end{eqnarray}

ii) for relatively strong field, $B \simeq 10 B_e \simeq 4 \times 10^{14}$ G,
the neutrino mean free path can be obtained from Eqs.~(\ref{eq:w_our_gen}), (\ref{eq:phy_small_eta}):
\begin{eqnarray}
\lambda \simeq \frac{3.2 \, \textrm{cm}}{B_{10}^{3/2} \, \sin \theta} \, 
\exp \left( \frac{4.0}{B_{10} \, E_{15}^2 \, \sin^2 \theta} \right) ,
\label{eq:lambda_s}
\end{eqnarray}
where $B_{10} = B/ (10 B_e)$, and the cutoff energy corresponding to 
$\lambda =$ 1 km, at $B_{10} = 1$, $\theta = \pi/2$, is 
\begin{eqnarray}
E_c \simeq 0.6 \times 10^{15} \textrm{eV} \,.
\label{eq:E_c_s}
\end{eqnarray}

The results obtained show an essential influence of the intense magnetic field 
on the process $\nu \to e^- W^+$ width. Despite the exponential character of suppression 
of the width in a strong field, Eqs.~(\ref{eq:w_our_gen}),~(\ref{eq:phy_small_eta}), as well as in a weak field, 
Eq.~(\ref{eq:w_Bor_small_chi}), the decay width in a strong field is greater in orders of 
magnitude than the one in a weak field, for the same neutrino energy. 

\section{Conclusion}

\begin{itemize}
\item
An influence of a strong external magnetic field on the neutrino self-energy operator is investigated. 
\item
The width of the neutrino decay into the electron and $W$ boson, and the mean free path of 
an ultra-high energy neutrino in a strong magnetic field are calculated. 
\item
An energy cutoff for neutrinos propagating in a strong field is defined. 
The cutoff energy at $B \simeq 5 B_e \simeq 2.2 \times 10^{14}$ G is: $E_c = 10^{15} \textrm{eV}$ .
\end{itemize}

\section*{Acknowledgements}

A.K. and N.M. express their deep gratitude to the organizers of the 
Seminar ``Quarks-2010'' for warm hospitality. We thank D.A. Rumyantsev for helpful remarks. 
 
This work was performed in the framework of realization of the Federal Target Program 
``Scientific and Pedagogic Personnel of the Innovation Russia'' for 2009 - 2013 (State contract no. P2323) 
and was supported in part by the Ministry of Education 
and Science of the Russian Federation under the Program 
``Development of the Scientific Potential of the Higher 
Education'' (project no. 2.1.1/510).



\end{document}